\documentclass[12pt]{iopart}

\usepackage{graphicx}

\begin{document}

\title{Mechanical probing of liquid foam aging.}
\author{I. Cantat\dag\ and O. Pitois\ddag  
\footnote[3]{isabelle.cantat@univ-rennes1.fr}
}

\address{\dag\ GMCM, UMR CNRS 6626, bat. 11a, campus de Beaulieu, 35042 Rennes cedex, France} 

\address{\ddag\ LPMDI, UMR CNRS 8108, 77454 Marne-la-Vall\'ee Cedex 2, France}

\begin{abstract}
We present experimental results on the Stokes experiment performed 
in a 3D dry liquid foam. The system is used as a rheometric tool :
from the force exerted on a 1cm glass bead, plunged at controlled velocity in 
the foam in a quasi static regime, local foam properties are probed around the sphere. With this original 
and simple technique, we show the possibility of measuring the foam shear modulus, the gravity drainage rate 
and the evolution of the bubble size during coarsening.
\end{abstract}

\maketitle

\section{Introduction}
Liquid foams belong to the large family of structured materials whose complex rheological properties are 
closely related to the organization at small scale. In the case of foams made of millimetric bubbles, this structure is directly observable \cite{lambert05} and ruled by relatively simple laws \cite{weaire}. This material is thus a very promising model system 
to understand the coupling between the disordered structure at small scale and the complex macroscopic rheological behavior
\cite{kraynik88}. 
Liquid foams behave like elasto-plastic materials and the force exerted on an obstacle moving in the foam at constant velocity
oscillates around a threshold value, which does not vanish at small obstacle velocity. The measure of this force is a very interesting probe for the foam behavior, and its averaged value has been measured in various regime in 2D \cite{dollet05} or in 3D  \cite{debruyn04,cox00}.
In these experiments, a sphere is moved through the foam and the drag force is measured as a function of the velocity. The system is thus used as a rheometric tool: using macroscopic constitutive relations the shear modulus of the foam is deduced from the data, although the rheometric flow is obviously more complex than in a conventional rheometer. However, one aspect of the method is of particular interest: the flow is confined in a sheared region surrounding the moving sphere. The force measurement is thus sensitive to the local foam properties.
In this paper, we focus on the measurement of foam properties using a high resolution mechanical probe, directly inspired from the Stokes experiment. We aim to demonstrate the possibility to deduce quantitative 
information on foam drainage, foam coarsening and on the elastic modulus from the measure of the force 
fluctuations on a moving sphere. 

\section{Experimental set-up}

\begin{figure}[h]
\centering
\begin{minipage}[c]{6cm}
\centering
\includegraphics[width=5.5cm]{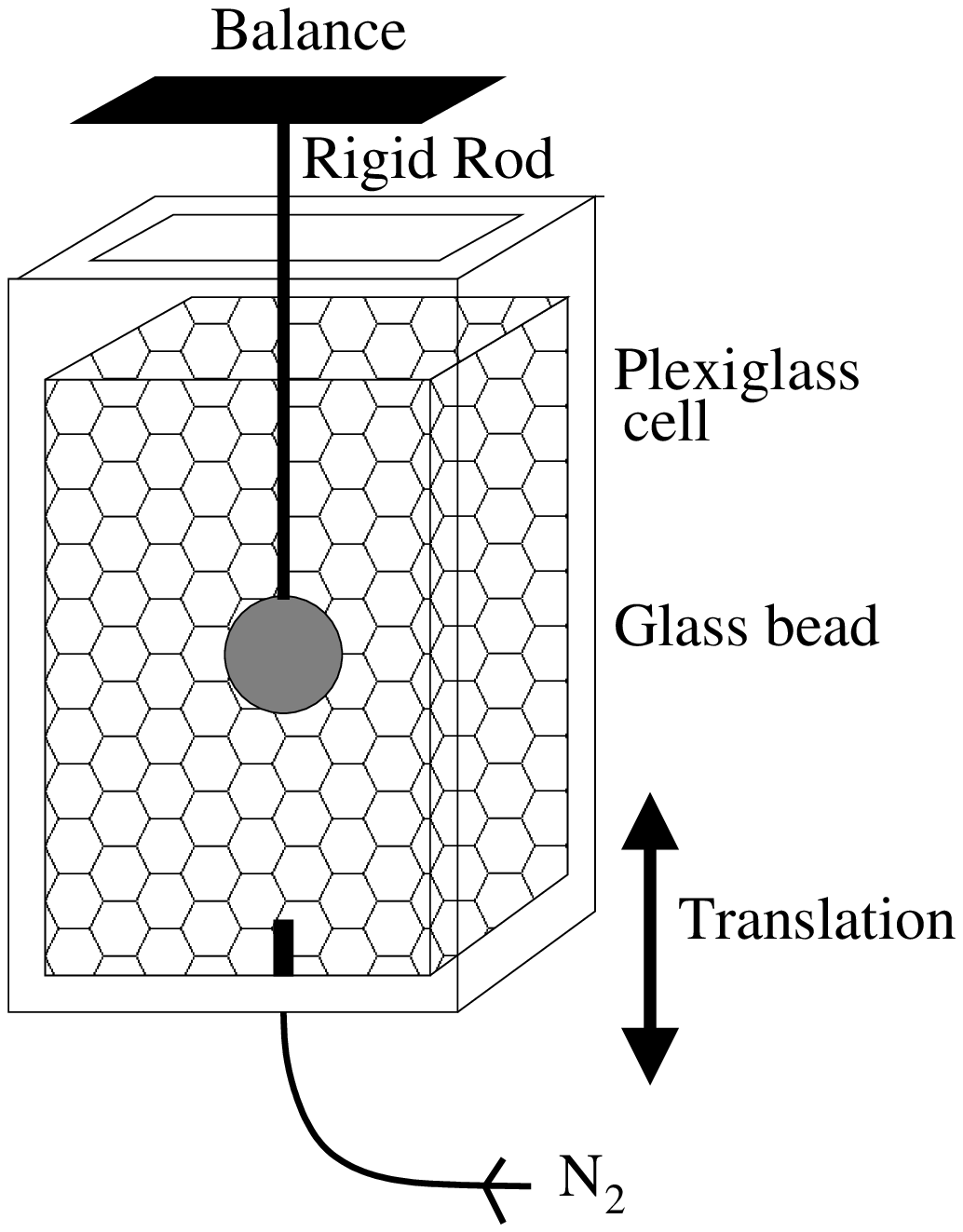}
\end{minipage}
\hspace{0.5cm}
\begin{minipage}[c]{6cm}
\centering
\includegraphics[width=6cm, angle=-90]{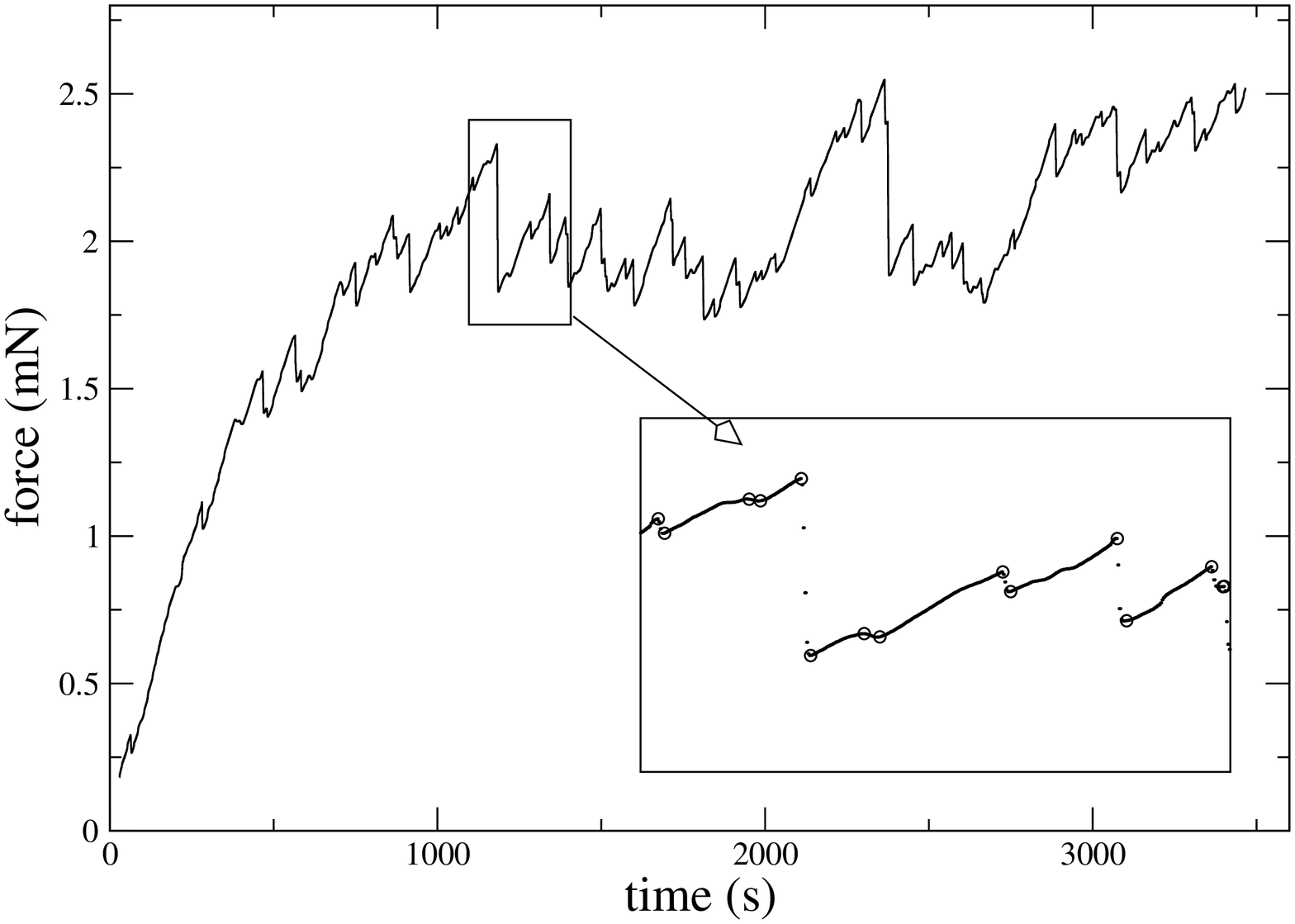}
\end{minipage}
\caption{\it (a) Experimental set-up. (b) Typical force graph obtained with the foam moving at 5 $\mu m/s$ downwards (positive forces are oriented downwards). The circle on the enlarged graph indicate the force extrema and the small dots the discrete measure points.}
\label{experience}
\end{figure}
The foam is produced in a $10 cm \times 10 cm \times 20 cm$ Plexiglas box by blowing nitrogen in a solution of SDS 3g/L and dodecanol 0.01g/L. The surface tension is $\gamma = 37 mN/m$.
The initial bubble size, measured by the mean bubble height in a corner of the box is of the order of few millimeters, with less than $10\%$ polydispersity.  The initial liquid fraction is estimated to be in the range 3-5 $\%$.
A glass bead of 1cm diameter, stuck to a  glass
rod of diameter 1mm and attached below a precision scale, is plunged in the foam.
The box can translate vertically 
at very low velocity ($ \pm 5 \mu m/s$) and thus imposes the foam quasi static motion around the immobile 
bead (see Fig.\ref{experience}(a)). 
The foam is not destroyed by the bead and, 
if the box is closed in order to avoid evaporation, the foam remains stable for hours. 
The force exerted by the foam on the bead is recorded every 0.33s by the scale with a precision of $1 \mu N$. The reference force is taken without foam in the box.
As we will show, the drainage occurs during the first half hour and the coarsening is significant after more than one hour. One-hour old foams are thus stable and have been used to calibrate the relation between the force and the foam properties. Younger and older foam has been used to study the drainage and the coarsening, respectively.

\section{Force calibration and shear modulus }

\begin{figure}[h]
\includegraphics[width=6cm, angle=-90]{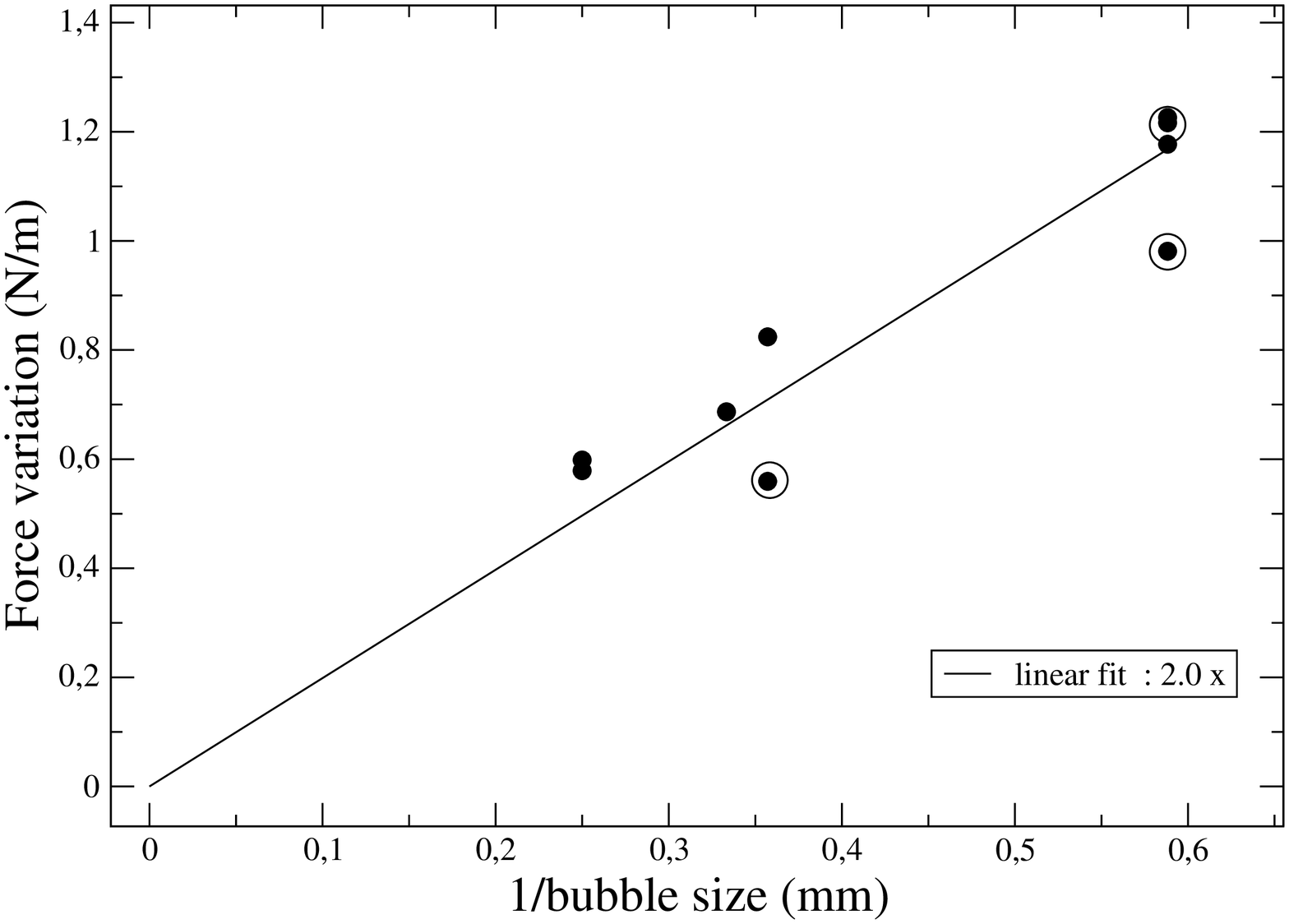}\hfill 
\includegraphics[width=6cm, angle=-90]{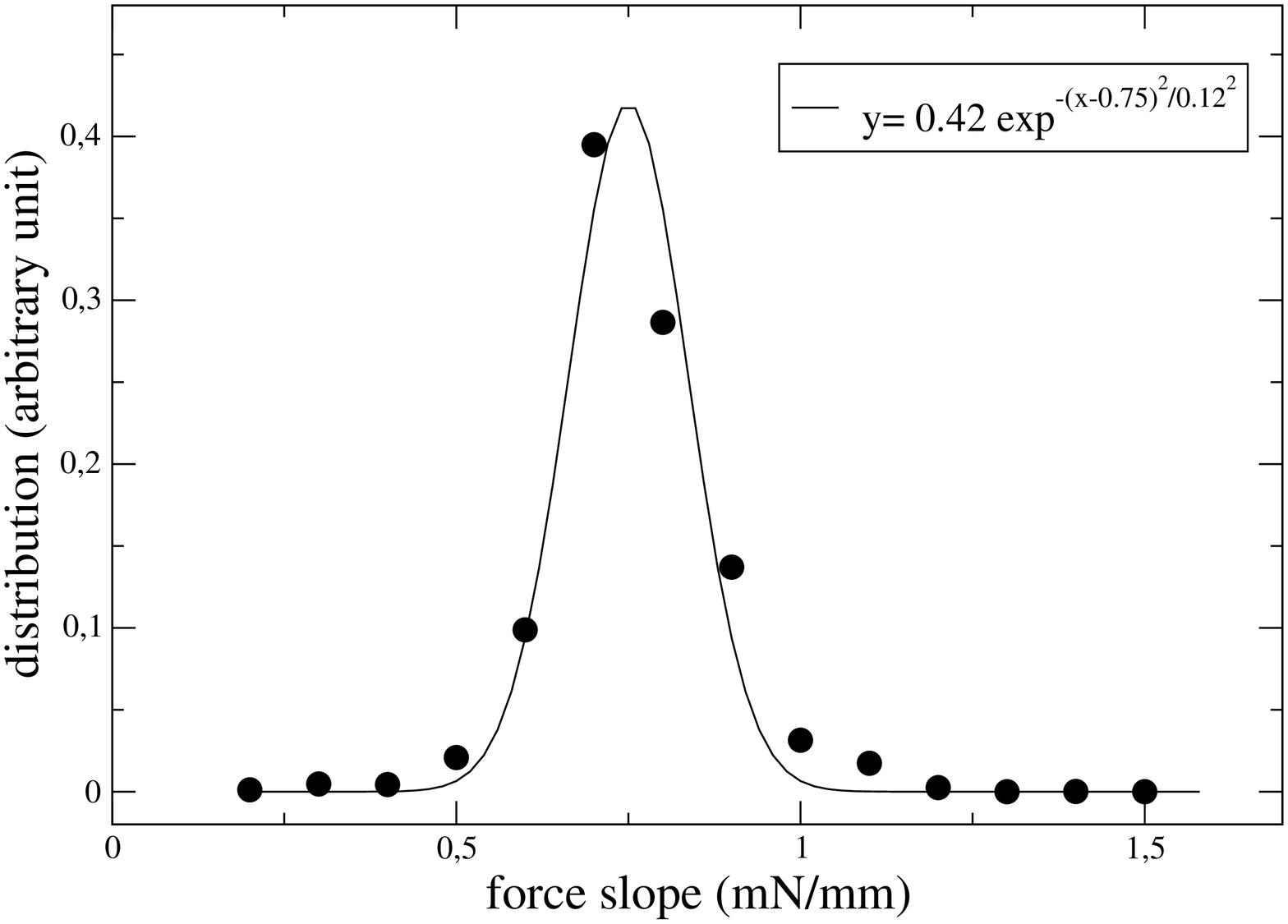}
\caption{\it (a) Force derivative during elastic loading with respect to the cell displacement, 
for various bubble sizes. The open circles indicate a box moving downwards. (b) Force slope distribution in a 3mm bubbles foam. This distribution, fitted by a Gaussian curve, is due to the experimental noise but also to the intrinsic local fluctuations of the foam properties.}
\label{pente_callib}
\end{figure}

If the box is moving downwards, the force exerted by the foam on the bead
is oriented  downwards as well and the scale signal is positive. During a transient 
the mean force increases until it reaches a threshold. Then the force signal becomes very similar to a stick slip signal, oscillating around the threshold value. Fig.\ref{experience}(b) evidences this succession of foam elastic loading, with a linear force increase, and of plastic rearrangements, corresponding to sudden force drops. This behavior has been predicted numerically \cite{weaire94} and observed in sheared bubbles raft \cite{lauridsen02} but, to our knowledge, 3D experimental measures are presented here for the first time.
The force signal is similar for both directions of motion and the  rod has no influence. 
In order to quantify the force increase rate, we only retain force drops larger than $\epsilon = 5 \mu N$. Then a force jump begins when $F(t+dt) < F(t) - \epsilon$ and finishes when $F(t+dt) > F(t) + \epsilon$. The opposite sign is taken if the foam moves upwards. From the list of force maximum $F_{max}$ occurring at $t_{max}$ and force minimum, $F_{min}$ at $t_{min}$, indicated by circles on Fig. \ref{experience}(b), we compute the rates of force
increase per unit time,  $dF/dt$, for each elastic loading.
These force slopes are very constant and reproducible. Even for very long elastic loading, the force increases remain linear. For foam velocities varied between $5 \mu m/$ and $50 \mu m /s$, the force graphs are identical if the time is rescaled according to each velocity, which evidences that the flow regime is quasi static. The force derivative with respect to the foam  displacement is thus independent on the velocity and simply verifies $V dF/dx = dF/dt$, with $V$ the cell velocity and $x$ the cell position. For various bubble sizes, we computed the slope distribution and the average slope value $<dF/dx>$
using a weight proportional to the elastic loading duration (see Fig. \ref{pente_callib}). 
We obtained a mean value of the slope which depends on the bubble size as $dF/dx \sim 1/d$ (see Fig.\ref{pente_callib}(a)). This shows good agreement 
with elastic predictions leading to $dF/dx = 4 \pi G R$, $G$ being the foam shear modulus and R the bead radius. This result
was obtained using an incompressible elastic medium with a  free sliding  condition on the bead, in an infinite cell. The medium 
is assumed to be at rest if the bead center is located at $x=0$. For a bead displacement of amplitude x, the elastic displacement at a distance $r$ from the bead  scales like $x \, R/r$, of the order of $0.1 x$ at the box boundary and, as shown by de Bruyn \cite{debruyn04}, the finite size effects are negligible even for smaller aspect ratio. 
A dimensional analysis implies that the foam shear modulus is $G = k \gamma/d$ for a foam, with $k$ an unknown number.
 This leads to $dF/dx = 4 \pi  R k \gamma/d$ or equivalently
\begin{equation}
dF/dt = 4 \pi  R V k \gamma /d \; . 
\label{dfdt}
\end{equation}
From the prefactor measured on Fig. \ref{pente_callib}, we deduce the foam shear modulus value
\begin{equation}
G = 0.85 {\gamma \over d} \pm 15\% \; . 
\label{Gded}
\end{equation}
The force slope distribution 
is relatively narrow, and well fitted by a Gaussian distribution. The error bar has been determined from the Gaussian width and is consistent with the points dispersion on Fig. \ref{pente_callib}(a).
In the literature, the shear modulus is usually expressed as a function of the radius $r_V$ of the 
sphere of same volume as the bubbles.
We measured the bubble radius in the solution during the foaming process and we found $r_V=0.54 \,d$.
This leads to the expression for the shear modulus G :
\begin{equation}
G = 0.46 {\gamma \over r_V} \pm 15\% \; . 
\end{equation}
A review of previous experimental and theoretical studies was done by Coughlin et al. (see \cite{coughlin96} and references therein). The values obtained for dry foams
in parallel plates or  Couette oscillating rheometers are around $G = [0.5 - 0.6 ]\; \gamma  / r_V$, in good agreement with theoretical models.  
Our result is slightly below these values. With a local and static solicitation (indentation test) Coughlin et al. found $G = 0.4 \, \gamma / r_V$.    The discrepancy between these measures may be due to experimental artifacts (wall slip ...), as suggested 
in \cite{coughlin96}, or due to more fundamental grounds as the influence of the geometry (local or not) and of the kind of flow used (oscillating with small amplitude, 
static, or steady flow). Numerical work performed by Reinelt and Kraynik \cite{reinelt96} simulated the
microrheological behavior of ordered foam samples. It was shown that these ordered
structures could exhibit highly anisotropic elastic responses. The Kelvin foam, for
example, have two shear moduli ($G_{min}=0.356 \; \gamma  / r_V $ and $G_{max}=0.6\; \gamma  / r_V$). As our foam sample is highly
monodisperse and as the flow is localized around the sphere, such anisotropic effect can
not be completely excluded. This important question will be the aim of a future work.

Once the prefactor is calibrated, the relation \ref{Gded} will allow to measure in a very simple way the bubble size
during the coarsening process (see section \ref{sec_coarsening}) and, for young foam with constant bubble size, the real relative velocity between 
bead and foam, which is influenced by gravity drainage
(see section \ref{sec_drainage}).

\section{Measurement  of liquid drainage.}
\label{sec_drainage}

\begin{figure}[h]
\centering
\includegraphics[width=6cm, angle=-90]{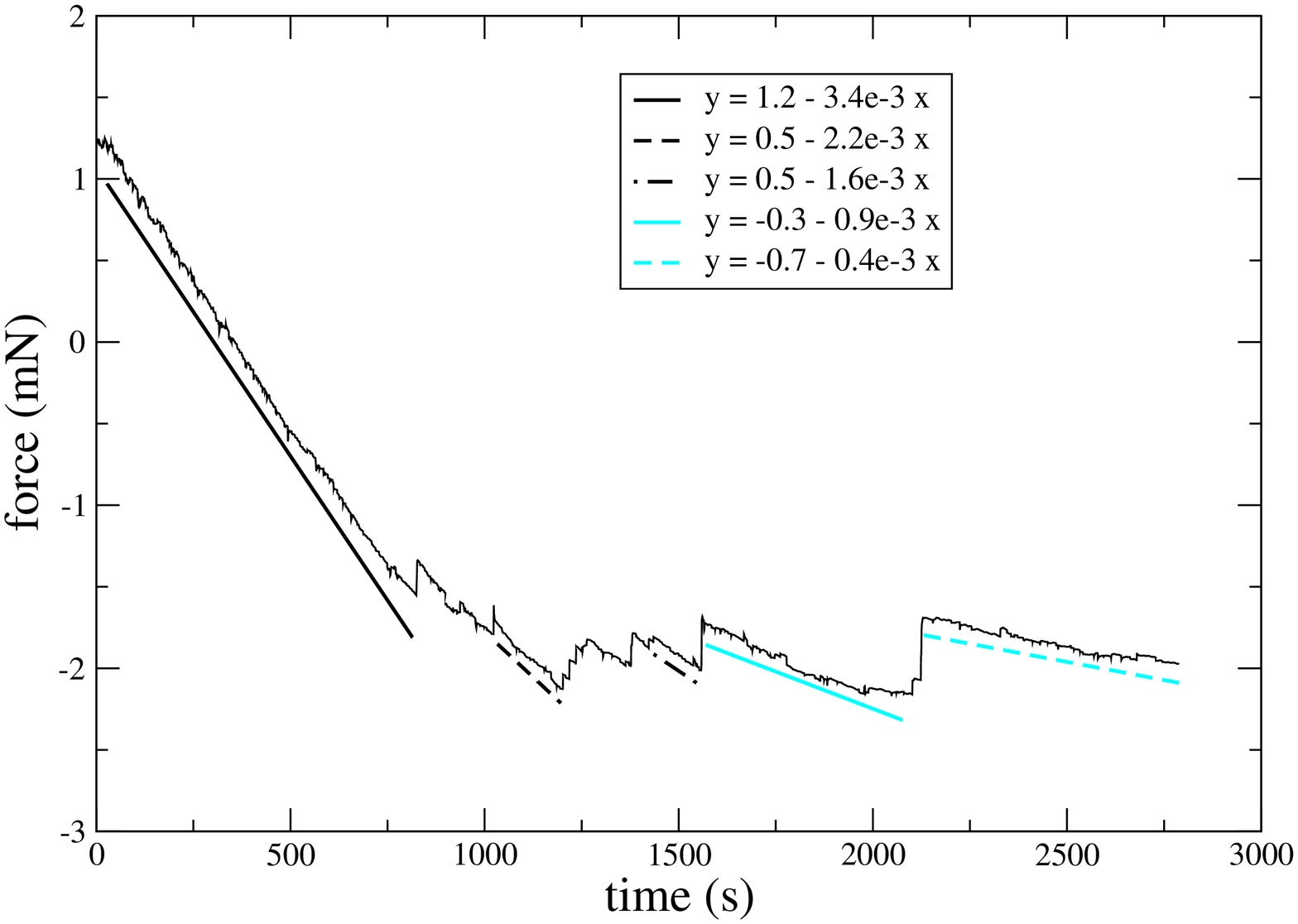}
\includegraphics[width=6cm, angle=-90]{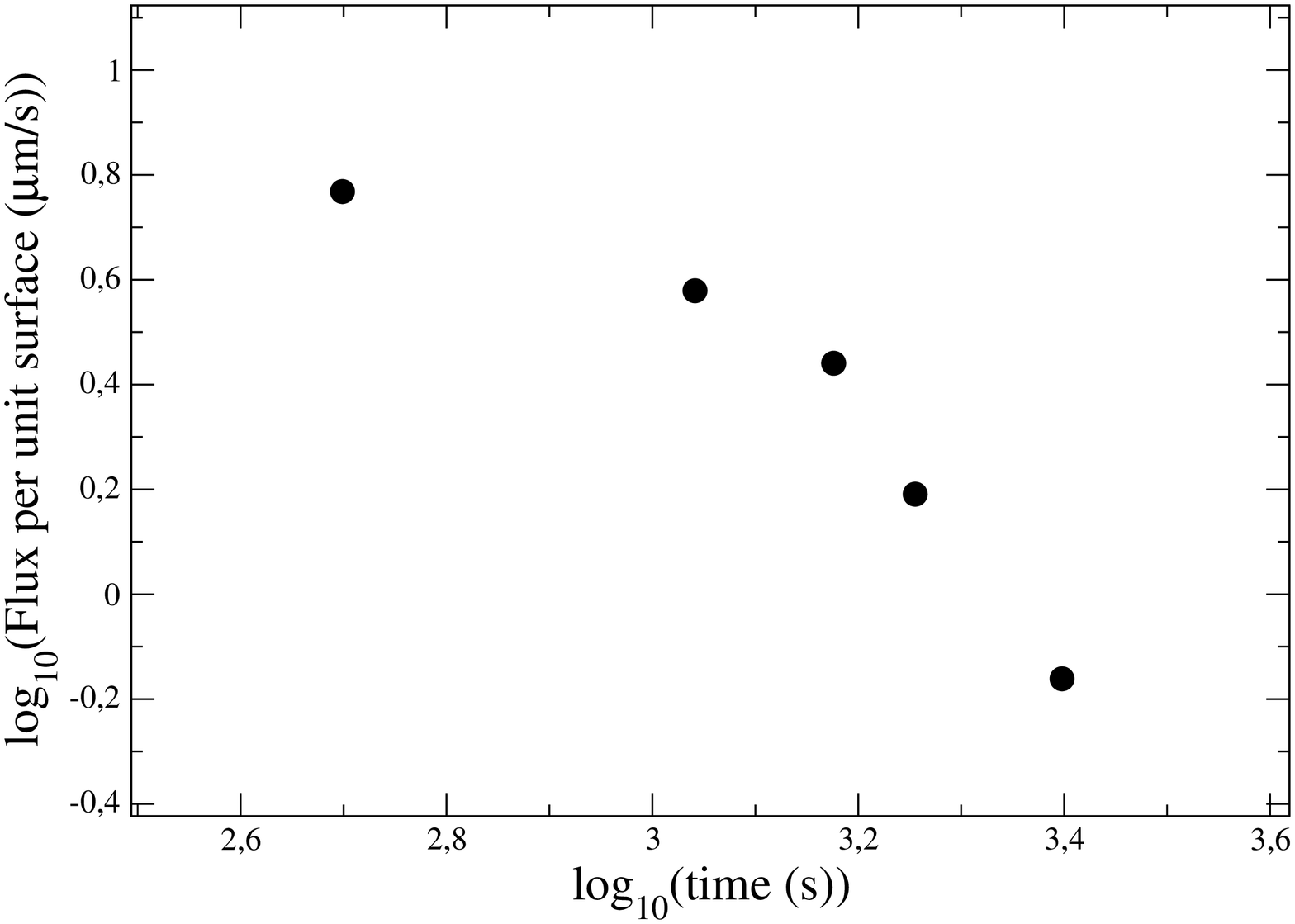}
\caption{\it (a) Force signal obtained for a young foam with bubble diameter 3 mm, when the bead is at rest. The slow force decreases are fitted by linear laws. (b) Evolution of the solution flux induced by gravity drainage, deduced using eq. \ref{dfdt} from the slopes of the linear fit given on graph (a). }
\label{drainage}
\end{figure}
Young foams are subjected to severe gravity drainage. Due to this outflow of liquid that accumulates at the bottom of the cell,  there is a net upward motion of the bubbles with a velocity $v_{foam}$. As a result, the relative velocity of the sphere through the foam is non-zero, even if the cell is not moving. Taking advantage of this superimposed velocity, a quantitative information on gravity drainage can be provided.
For a given horizontal cross-section $A$ of the cell, the upward flow rate of the foam $q_{foam} =A v_{foam}$ exactly scales the gravity flow of liquid through the Plateau borders $q_{Pb}= N_{Pb} a_{Pb} \bar{u}$, where  $N_{Pb}\sim A/L^2$
 is the number of PB channels in the cell cross-section, $L$ the PB length, $a_{Pb}$ is the PB cross-section area and $\bar{u}$ is the mean liquid velocity through the Plateau borders.  In the limit of a dry foam, $a_{Pb}$ can be related to the foam liquid fraction $\epsilon$: $a_{Pb}\sim \epsilon L^2$, so that $v_{foam}\sim -\epsilon \bar{u}$. Expressions for $\bar{u}$ can be found in the literature. The study of the drainage behavior is beside the scope of this paper, and we prefer, for the sake of simplicity,  to follow the simplest approach, i.e. a Poiseuille flow through the Plateau borders. Neglecting capillary forces we thus obtain $\bar{u} \simeq  (\rho g L^2/\mu) \epsilon$ \cite{verbist96}, where $\mu$ is the bulk shear viscosity of the liquid solution. This approach is also justified by the presence of dodecanol in the solution, which is known to be responsible for the rigidification of the interface. The upward foam velocity is thus related to the liquid fraction by  $v_{foam} \sim (\rho g L^2/\mu) \epsilon^2$.
From the graph force presented in figure \ref{drainage}(b), we measure $dF/dt$ during elastic loading and the foam velocity is obtained  from eq.\ref{dfdt}.  We deduce the variation of the liquid fraction in the horizontal plane of the sphere $\epsilon \sim t^{-\alpha}$, with $0.6 < \alpha < 1$. This result is consistent with theoretical predictions for free drainage, in the case of rigid interfaces \cite{koehler00,saintjalmes02}.
\begin{figure}[h]
\centering
\includegraphics[width=6cm, angle=-90]{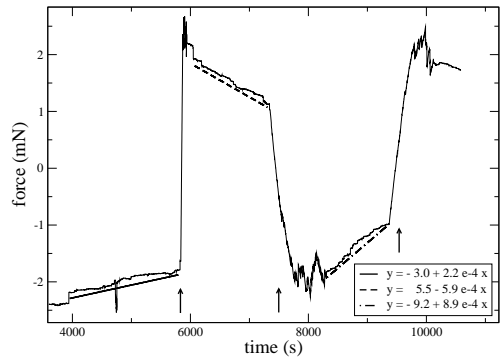}
\caption{\it Force evolution for an old foam. The bead is at rest, excepted for the three modifications of the foam loading, indicated by arrows. Between these small bead translations the foam creeping motion is observed. }
\label{creep}
\end{figure}

This phenomenon is observed independently of 
the initial loading of the foam, thus if the foam is initially moved upwards (because of the production process for example)
the absolute value of the force exerted on the bead increases with the time (see Fig. \ref{drainage}(a)). It can therefore not be mistaken for a creeping motion, obtained at longer time in the same experimental conditions. In this creeping regime, if the bead is alternatively moved in the foam and then left at rest (see Fig \ref{creep}), we observe that 
the force relaxation is faster for older foam, which can be explained by a larger noise induced by coarsening \cite{cohenaddad04}.

This technique allows for direct measurement of the solution flux under the sole assumption that the foam container is closed at the bottom and that the solution flux is homogeneous over the whole horizontal section passing trough the bead. It is thus complementary with other methods that usually measure the liquid fraction. The prefactor $k$ in eq.\ref{dfdt} may be obtained from the same foam, at the end of the drainage process (which we did), or in the general case, 
by comparison between the slope obtained with a cell at rest and with a cell moving at a controlled velocity. 
At each time, the foam velocity and the local shear modulus can thus be determined simultaneously. 
In case of high flux forced drainage, convection appears in the cell and the gas and liquid fluxes vary horizontally \cite{weaire03}. In this case, our technique would be very useful too, not to obtain the solution flux but directly to measure the local foam velocity.

\section{Coarsening measure.}
 \label{sec_coarsening}
 \begin{figure}[h]
\centering
\includegraphics[width=6cm, angle=-90]{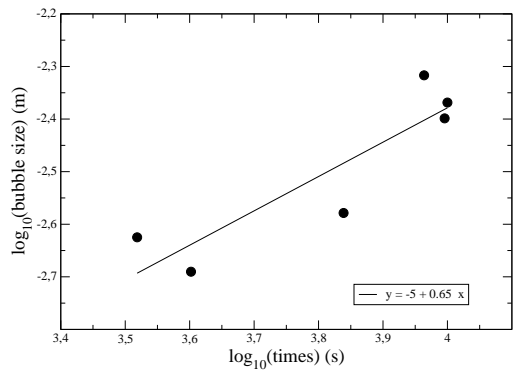}
\caption{\it Evolution of the bubble diameter with time. Despite the small parameter range, a logarithmic plot is displayed with a slope compatible with the power law prediction $d\sim t^{0.5}$ in scale invariant coarsening regime. }
\label{coarsening}
\end{figure}

For drained foam, after a creeping motion transient, the force signal  vanishes if the cell does not move. The relative velocity between foam and bead is then simply controlled by the cell motion, which is fixed at $5 \mu/s$. Three foams with the same initial bubble 
size (of the order of 2mm) were followed during several hours, with upwards or downwards cell motion. 
The foam is not broken when the bead is driven through the medium and the force response is the same when the bead passes for the first or second time, for foams of same age. The direction of motion does not modify the force. The bubble size is deduces from eq. \ref{dfdt} and plotted on Fig. \ref{coarsening}. 
We obtain that the typical bubble diameter evolves as $d \sim t^{0.65}$, which is compatible with the exponent $0.5$ expected
for dry foams \cite{glazier00}.

\section{Conclusion}
In this paper, we propose a new and simple technique to measure local foam properties, based on 
force measurements on a spherical obstacle moved at a controlled velocity in the foam. From force increase rate between two successive plastic events, we deduce
the foam shear modulus, related to the local bubble size, and the relative velocity of the foam. The variation of the liquid fraction
during gravity drainage, as well as the evolution of the  mean bubble size during coarsening process was successfully measured with this technique. One of the most promising applications is probably to determine local velocity fields during quasi static motions.
The typical diameter of the foam volume probed by the bead remains to be measured, in order to estimate the spatial resolution of this measure.

\vspace*{0.5cm}

{\bf Acknowledgments}\\
The authors are grateful to the French Spatial Agency (CNES) and to the CNRS for financial support. We thank Mr Hautemayou and Mr Laurent for technical help. IC thanks the LPMDI for its hospitality.

\end{document}